\documentclass[twocolumn]{revtex4}
\usepackage{enumerate}
\usepackage{graphicx}
\usepackage{dcolumn}
\usepackage{amsmath}
\usepackage{epstopdf}

\usepackage{bm}

\makeatletter
\def\btt#1{\texttt{\@backslashchar#1}}%
\DeclareRobustCommand\bblash{\btt{\@backslashchar}}%
\makeatother
\begin{document}
\preprint{HEP/123-qed}
\title{ Studies on binary mixtures of nematic liquid crystals made of strongly polar molecules with identical cores and antagonistic orientation of permanent dipoles}

\author{Dasari Venkata Sai$^{1,2}$,  Tae Hoon Yoon$^{1*}$, and Surajit Dhara$^{2}$}
\email{thyoon@pusan.ac.kr; sdsp@uohyd.ernet.in} 

\affiliation{$^1$Department of Electronics Engineering, Pusan National University, Busan 609-735, Korea\\
$^2$School of Physics, University of Hyderabad, Hyderabad-500046, India}


\date{\today}
\begin{abstract}
We report experimental studies on optical (birefringence, $\Delta n$), dielectric $(\Delta \varepsilon)$ and bend-splay elastic anisotropies ($\Delta K=K_{33}-K_{11})$  of a few mixtures of two nematic liquid crystals, namely CCH-7 and CCN-47, made of highly polar molecules with identical cores and antagonistic orientation of permanent dipoles.  In particular, the polar group (-CN) attached to the bicyclohexane core of CCH-7  is oriented along the longitudinal direction whereas, in CCN-47, it is oriented along the transverse direction.  We show that apart from the significant contribution to the optical and dielectric anisotropies, the antagonistic orientation of strongly polar groups plays a crucial role in determining the bend-splay elastic anisotropy. The elastic properties  are explained based on a model proposed by Priest, considering the effect of intermolecular association and the resulting length-to-width ratio of the molecules.

\end{abstract}
\preprint{HEP/123-qed} 
\maketitle

\section{ Introduction}

 The electrooptical response of liquid crystals (LCs) mostly depends on the birefringence, dielectric, elastic properties, and the rotational viscosity~\cite{de,ch,deju,blinov}. Depending  upon the applications, a set of desired physical properties is required. Since all the desired properties are  unavailable in any single compound, usually several liquid crystals are mixed in appropriate proportions to tune the physical properties for commercial devices.  Further, studies on  liquid crystal mixtures with a variety of shapes and structures have always proven to be rewarding from both the technological and fundamental perspectives. Several new phases have been discovered, which are absent in the parent compounds~\cite{pa,sd1,rp}. In the context of binary mixtures made of symmetric and highly asymmetric molecules, such as mixtures of rod-like and bent-shaped molecules have resulted in several new physical properties \cite{har,kundu,satya,satya1,dodge,dodge1}. However, physical studies on the binary mixtures of nematic liquid crystals with the identical core structures and highly antagonistic dipole orientation are meagre. High dipole moments of the molecules are known to make an important contribution to the intermolecular interactions and give rise to distinct physical properties of liquid crystals~\cite{nvm1,nvm2,nvm3,nvm4,sd}.
 
\begin{figure}[htp]
\centering
\includegraphics[scale=0.8]{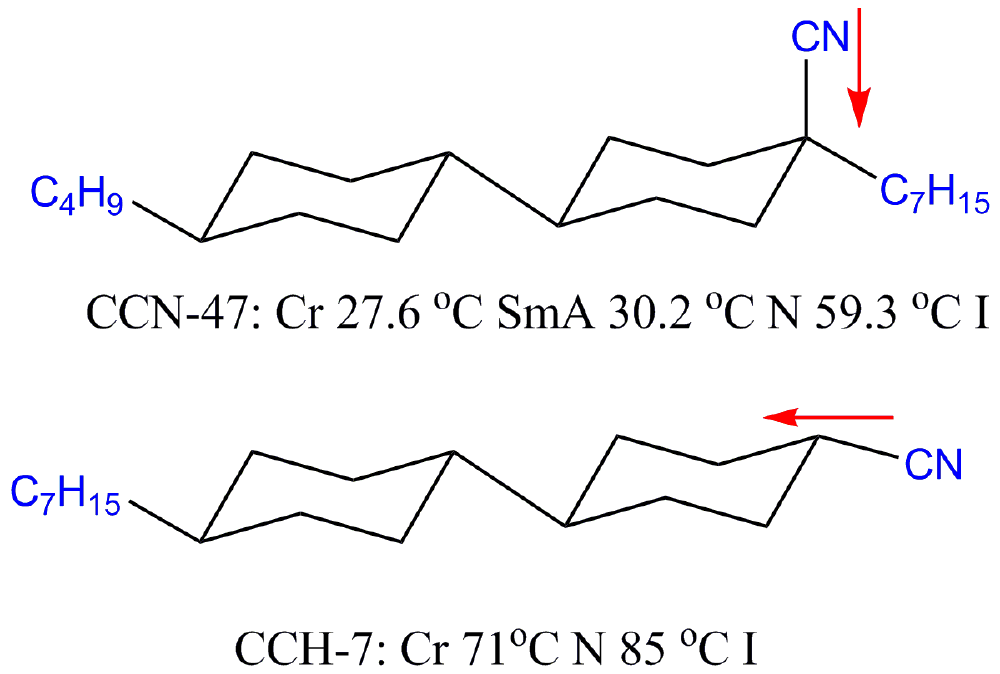}
\caption{ Chemical structures and phase transition temperatures of the liquid crystals. Red arrows indicate the direction of permanent dipole moment.}
\label{fig:figure1}
\end{figure}

 In this paper, we report experimental studies on the binary mixtures of two low molecular weight nematic liquid crystals, namely CCH-7 and CCN-47, having identical (bicyclohexane) cores and antagonistic dipole orientation with respect to their long axes (see Fig.\ref{fig:figure1}). In particular, the direction of permanent dipole moment  is oriented parallel and perpendicular to the long axis  in case of CCH-7 and CCN-47, respectively \cite{sai,gj}.   We have prepared a few mixtures with varying composition and measured physical properties such as birefringence, dielectric and elastic anisotropies as a function of temperature. We show that the orientation of the polar group (-CN) not only contributes to the optical and dielectric anisotropies, but it also significantly affects the bend-splay elastic anisotropy.

\section{Experimental} 

\begin{figure*}[htp]
\includegraphics[scale=0.7]{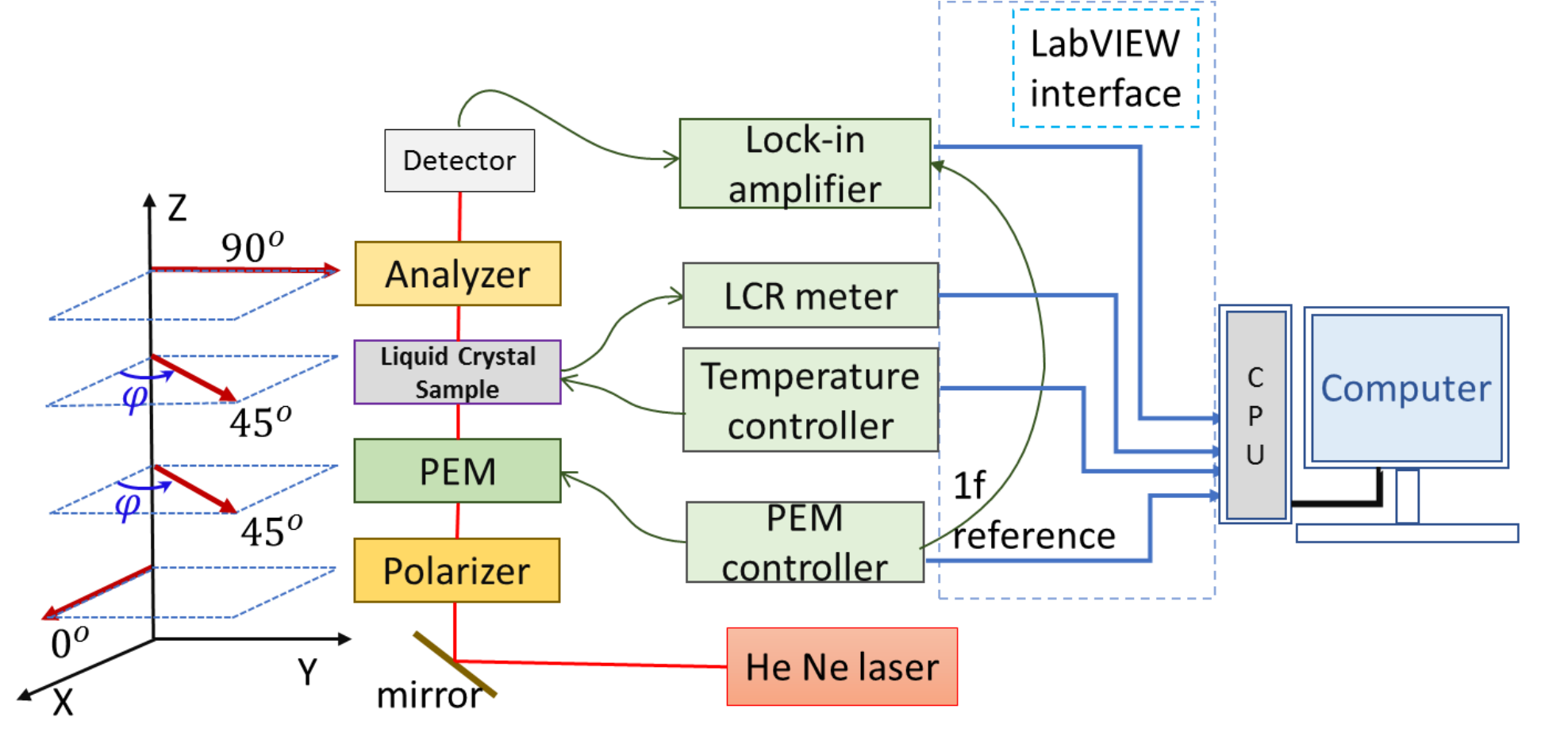}
\caption{ Schematic diagram of the experimental setup. The orientations of the polariser, analyser, photoelastic modulator (PEM) axes and the director  are shown by red arrows.}
\label{fig:figure2}
\end{figure*}
The experimental cells are made of two parallel indium-tin-oxide coated glass plates with circularly patterned electrodes. To obtain the planar alignment of the director (the mean molecular orientation), the plates are coated with polyimide AL-1254 and cured at 180$^\circ$C for 1 hour. The polyimide coated plates are rubbed in an antiparallel way for homogeneous or planar alignment of the director. For  homeotropic alignment, the plates are coated with polyimide JALS-204 and cured at 200$^\circ$C for 1 hour. The typical cell thickness used in the experiments is about 8$\mu$m.  The  phase transition temperature of the mixtures are obtained using a polarising optical microscope (Nikon, LV100 POL)  and a temperature controller (Instec, mK1000). The temperature-dependent birefringence of the mixtures is measured by a polarisation modulation technique, using a home-built electro-optic setup, involving a He-Ne laser ($\lambda=632.8$nm), photo-elastic modulator (PEM) and a lock-in amplifier~\cite{oak,sat1,sat2}. A schematic diagram of the experimental setup is shown in Fig.\ref{fig:figure2}. Planar cells are used for all the experimental measurements on samples with positive dielectric anisotropy.  We measured voltage-dependent dielectric constant of the samples form 0.02 to 18V in steps of 0.03V, using an LCR meter (Agilent E4980A) at a frequency of 4111 Hz. For samples with positive dielectric anisotropy the dielectric constant below the Freedericksz's threshold voltage \cite{freedericksz} is equal to $\epsilon_{\perp}$. To measure $\epsilon_{||}$, the linear part of dielectric constant at higher voltages plotted against $1/V$ and extrapolated to 0 ($ V\rightarrow\infty \Rightarrow 1/V\rightarrow0$). This experiment is repeated at different temperatures to get temperature dependent dielectric anisotropy, $\Delta \epsilon$. A computer controlled LabVIEW program is used to control the experiments. \\
The same procedure is followed for the samples with negative dielectric anisotropy, but in homeotropic cells~\cite{sai}.  The splay and bend elastic constants of all the samples are measured following the procedure described in ref.\cite{gruler, morris,sai}. The birefringence and dielectric data are measured with an accuracy of 2\%. The splay and bend elastic constants of  positive dielectric anisotropy materials are measured with an accuracy of 5\% and 8\%, respectively. The bend and splay elastic constants of negative dielectric anisotropy materials are measured with an accuracy of  5\% and 10\%, respectively.

\section{Results and discussion}
The chemical structure of the pristine compounds are shown in Fig.\ref{fig:figure1} and the phase transition temperatures of the compounds are presented in Table-I. Apart from nematic, CCN-47 also exhibits a nematic to smectic-A (N-SmA) phase transition at 30.2$^\circ$C ($T_{NS}$). The molecules of both CCN-47 and CCH-7 have identical bicyclohexane cores. The polar group (-CN) in CCN-47 is directed along the transverse direction and exhibits negative dielectric anisotropy, whereas in CCH-7, the -CN group is directed along the longitudinal direction and exhibits positive dielectric anisotropy \cite{ananth, ibra,sai}. 
\begin{table}
\caption{ Phase transition temperatures of  binary mixtures  of CCN-47 and CCH-7. $T_{NI}$: nematic to isotropic; $T_{NS}$: N-SmA transition temperatures.  `-' indicates no N-SmA transition is observed in the experimental temperature range. }
\vspace{0.1in}
\begin{tabular}{|c|c| c| c| c|}
\hline
  & wt$\%$ CCN-47  & wt$\%$ CCH-7  & $T_{NI}$ ($^\circ$C) & $T_{NS}$ ($^\circ$C)  \\

\hline
Sample-1   &100   &   0   & 59.6 & 30.2 \\ 
\hline
Sample-2    &75  &  25   &  46.5 &  -   \\
\hline
Sample-3    & 55   &   45   &  47.6 &  39.2 \\
\hline
Sample-4    & 23  &   77   &  71.5 & -   \\
\hline
Sample-5    & 0   &  100   &  85.0  & -   \\
\hline
\end{tabular}
\end{table}

\begin{figure}[htp]
\centering
\includegraphics[scale=0.58]{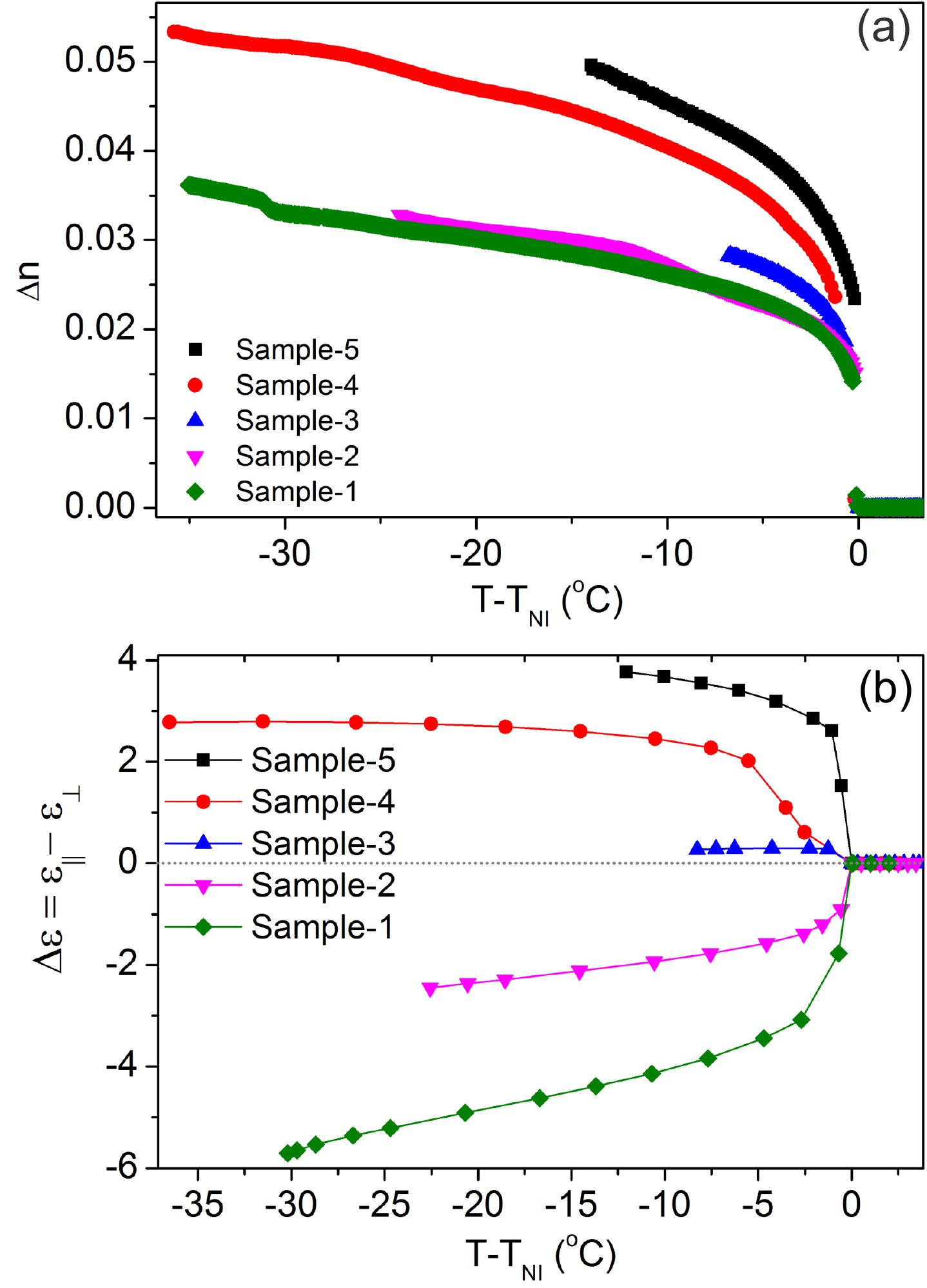}
\caption{ Variation of (a) birefringence $(\Delta n)$ and (b) dielectric anisotropy ($\Delta\epsilon$) of the mixtures as a function of shifted temperature.}
\label{fig:figure3}
\end{figure} 

We prepared three mixtures with different wt\% of these compounds as shown in Table-I.  Sample-1 and Sample-5 are pure  CCN-47  and CCH-7 compounds. All the mixtures show the nematic phase. Sample-3 shows relatively a shorter temperature range of nematic ($8.4^{\circ}$C) and a nematic to smectic-A phase transition.
The birefringence ($\Delta n$) of the samples  as a function of shifted temperature is shown in Fig.\ref{fig:figure3}(a). At a shifted temperature $T-T_{NI}=-12^{\circ}$C, the birefringence of pristine  CCN-47  is  $\Delta n\simeq0.027$ and for  pristine CCH-7, it  is $\Delta n\simeq0.047$ and agrees well with the previous reports  \cite{sai,ibra,ananth}. At a fixed shifted temperature, $\Delta n$ increases with increasing wt\% of CCH-7 in the mixtures as expected. Since the molecules have identical core structure, the difference in the birefringence (i.e., $\Delta n_{CCH-7}-\Delta n_{CCN-47}\simeq0.02$)  arises mostly due to the antagonistic orientation of the  -CN groups. 
The variation of dielectric anisotropy ($\Delta \varepsilon$) of the samples is shown as a function of shifted temperature in Fig.\ref{fig:figure3}(b). The dielectric anisotropy of pristine CCH-7 is a large positive ($\Delta\epsilon\simeq 3.8$), whereas for CCN-47, $\Delta \varepsilon$ is  a large negative ($\Delta\epsilon\simeq -4.2$) at $T-T_{NI}=-12^\circ$C.   The core structures of the two molecules are identical, hence the large difference in $\Delta \varepsilon$ of the pristine samples is mostly due to the difference in the orientation of the  -CN groups. This indicates that the antagonistic orientation of the -CN groups in mixture tends to cancel out their contribution on the dielectric anisotropy.
In fact, $\Delta \varepsilon$ of Sample-3, which is composed of nearly 55wt\% of CCN-47 is very close to zero ($\Delta\epsilon\simeq 0.03$). 

\begin{figure}[htp]
\includegraphics[scale=0.75]{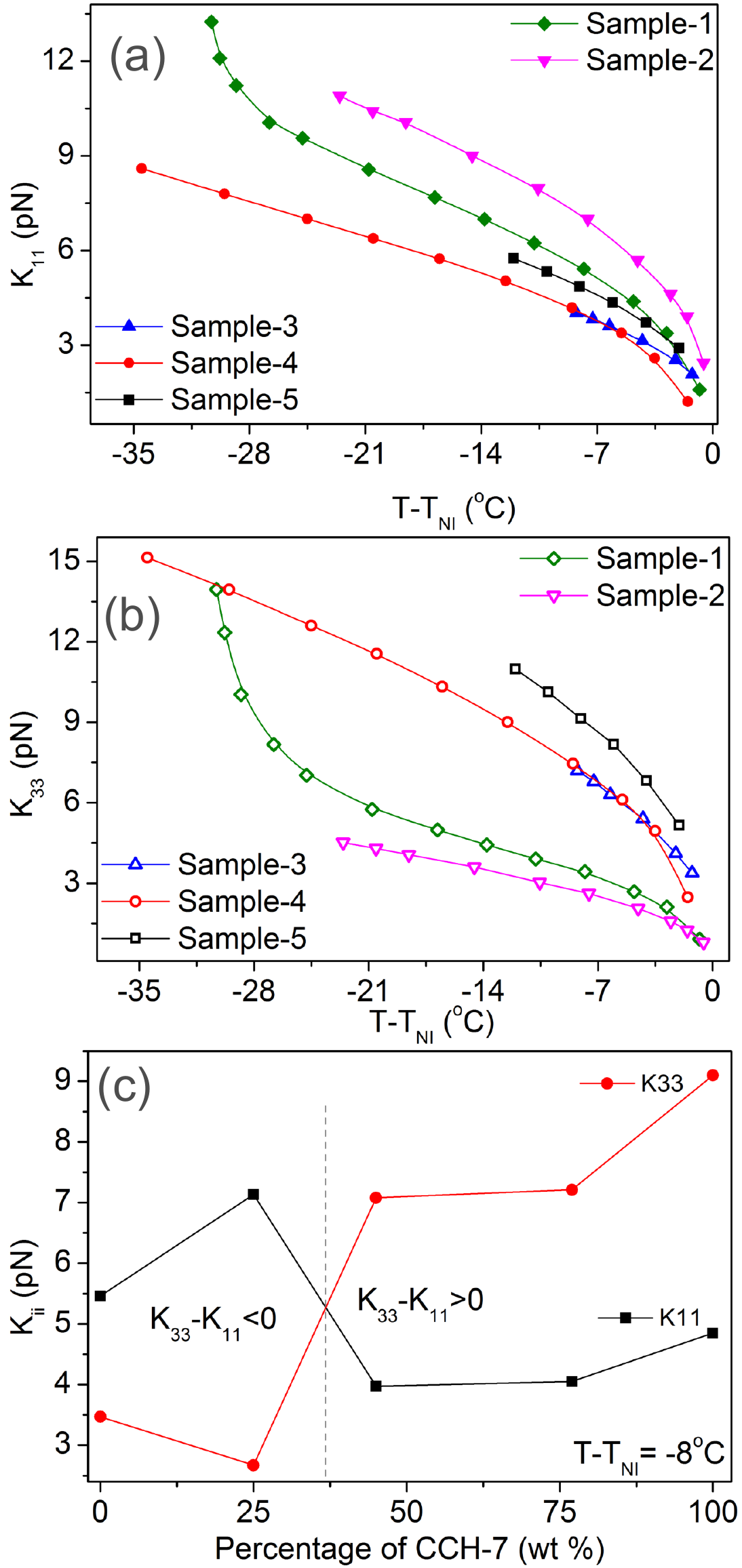}
\caption{ Variation of (a) splay ($K_{11}$)  and (b) bend ($K_{33}$)  elastic constants with shifted temperature. (c) Variation of $K_{33}$ and $K_{11}$ with wt\% of CCH-7 at a relative shifted temperature $T-T_{NI}=-8^\circ$C.}
\label{fig:figure4}
\end{figure}

\begin{figure}[htp]
\includegraphics[scale=0.7]{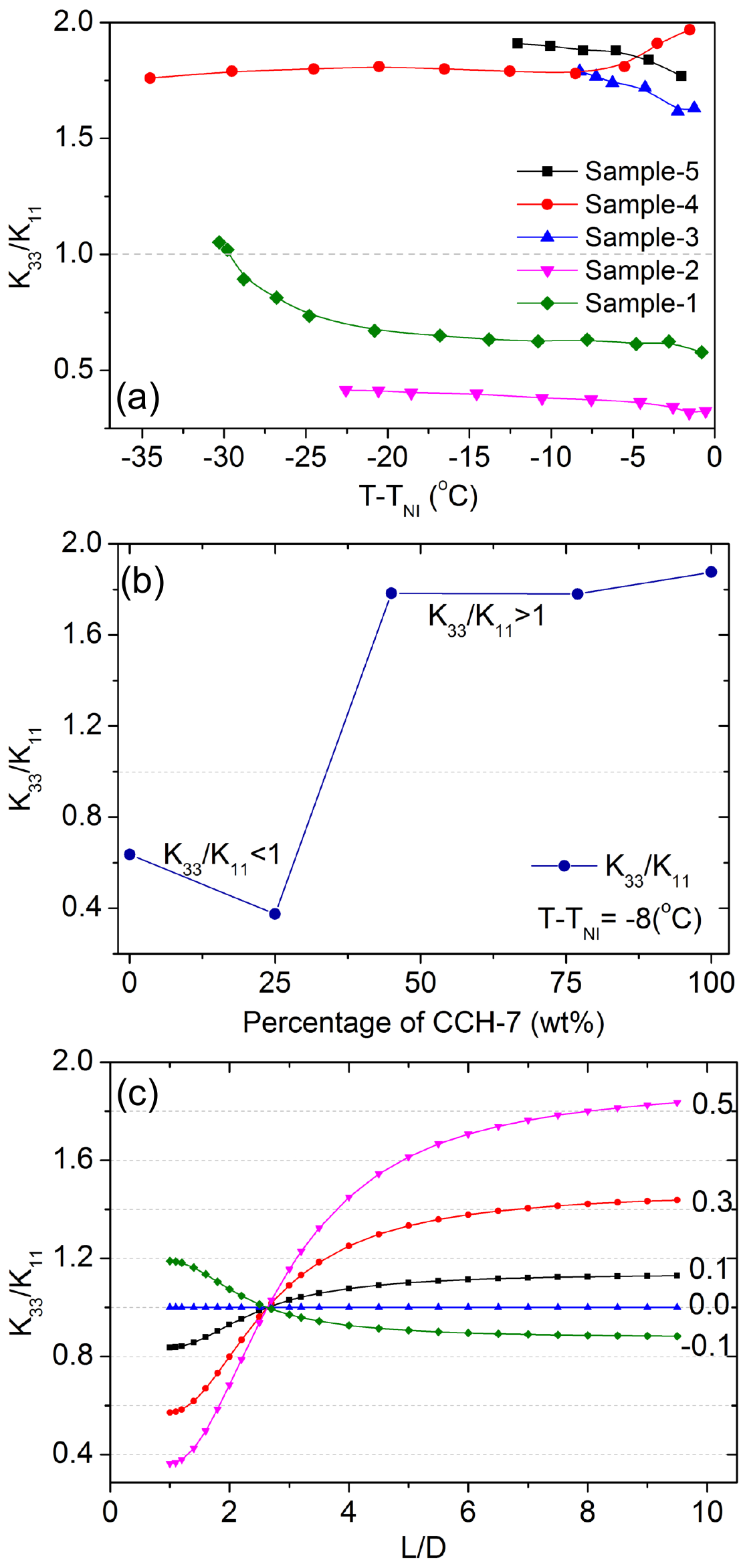}
\caption{ (a) Ratio $K_{33}/K_{11}$  (a) as a function of shifted temperature (b) as a function of wt\% of CCH-7 at a relative shifted  temperature $T-T_{NI}=-8^\circ$C. (c) Variation of calculated $K_{33}/K_{11}$ with length-to-width ratio ($L/D$) using Eq.(1) at a few values of $\overline{P_{4}}/\overline{P_{2}}$.}
\label{fig:figure5}
\end{figure}

The variations of splay $(K_{11})$ and bend $(K_{33})$ elastic constants of the samples as a function of shifted temperature are shown in Fig.\ref{fig:figure4}(a,b).  Both $K_{11}$ and $K_{33}$ increase with decreasing temperature as they are proportional to the square of the orientational order parameter $S$~\cite{de}. In pristine CCN-47 (Sample-1) both $K_{11}$ and $K_{33}$ tend to diverge near the room temperature due to the short-range presmectic order effect \cite{sai, sai1}. Figure~\ref{fig:figure4}(c) shows that in pristine CCN-47, the bend-splay elastic anisotropy ($\Delta K=K_{33}-K_{11}$) is negative whereas, in pristine CCH-7,  $\Delta K$ is positive. It changes sign between 25wt\% to 50wt\% of CCH-7.
Figure~\ref{fig:figure5}(a) shows the ratio of the two elastic constants ($K_{33}/K_{11}$) of the samples  as a function of shifted temperature. It is observed that for Sample-1 and Sample-2, $K_{33}/K_{11}<1$, whereas for Sample-3, Sample-4 and Sample-5,  $K_{33}/K_{11}>1$.  Figure~\ref{fig:figure5}(b) shows that at a fixed shifted temperature ($T-T_{NI}=-8^\circ$C) the ratio, $K_{33}/K_{11}$ increases with increasing wt\% of CCH-7.

The variation of $K_{33}/K_{11}$ can be qualitatively explained based on a molecular theory proposed by Priest~\cite{prist}, considering the effective length-to-width ratio of the molecules. The ratio $K_{33}/K_{11}$ is related to the molecular properties and given by \cite{prist}:
\begin{equation}
\frac{K_{33}}{K_{11}}=\frac{1+\Delta+4\Delta^{'}\overline{P_{4}}/\overline{P_{2}}}{1+\Delta-3\Delta^{'}\overline{P_{4}}/\overline{P_{2}}}
\end{equation}
 where $\Delta=(2R^2-2)/(7R^2+20)$, $\Delta^{'}=9(3R^2-8)/16(7R^2+20)$ and $R=(L-D)/D$. $L$ and $D$ are the length and width of the spherocylindrical molecules. The calculated ratio of $K_{33}/K_{11}$ for various values of $\overline{P_{4}}/\overline{P_{2}}$ is shown in Fig.\ref{fig:figure5}(c).  It is observed that  the ratio $K_{33}/K_{11}$ increases or decreases depending on the sign of $\overline{P_{4}}/\overline{P_{2}}$. For positive values of $\overline{P_{4}}/\overline{P_{2}}$,  the ratio $K_{33}/K_{11}$ increase with $L/D$.

 In the x-ray measurements of nematic phase of  CCH-7, an antiparallel local ordering of the molecules was reported  due to the antiparallel correlation of the longitudinal polar group (-CN) \cite{gj}. Bradshaw \textit{et. al.}, showed  the antiparallel local ordering of the dipoles increases the effective length of CCH-7 molecules \cite{brad}. As a result of which the ratio $K_{33}/K_{11}$ is greater than 1 (see Fig.\ref{fig:figure5}(a)).  In our previous study,  from the energy minimised DFT calculations, the shape of the CCN-47 molecule was found to be bent-shape \cite{sai}. Due to such a shape, the effective length-to-width ratio could be relatively smaller.  In addition, they do not exhibit strong antiparallel correlation due to the transverse orientation of the polar group (-CN)~\cite{praveen}. This could results in, $K_{33}/K_{11}<1$ for pure CCN-47. As we increase the wt\% of CCH-7, beyond about 25wt\%, the antiparallel correlation of permanent longitudinal dipoles (-CN) of CCH-7 develops, as a result of which the length-to-width ratio is increased. Thus, increasing wt\% of CCH-7 in the mixture is an equivalent effect of increasing molecular length. The best comparison to the experimental data with calculation is obtained with a length-to-width ratio of $L/D\simeq 7$ and a value of the ratio $\overline{P_{4}}/\overline{P_{2}}=0.5$. In the x-ray diffraction studies of CCH-7, the effective $L/D$ was reported to be $\simeq6.5$~\cite{gj}. This is very close to the value at which our experimental result resembles with the calculations (see Fig.\ref{fig:figure5}(c)). It would be very useful to have experimental measurements  of $\overline{P_{4}}/\overline{P_{2}}$ of these samples. It may be mentioned that both the rotational and translational entropy of the molecules with longitudinal dipole moment is expected to get reduced due to the antiparallel correlation. This effect should be absent in molecules with transverse dipole moment due to the lack of antiparallel correlation. As a result of which, the orientational order parameter  and eventually the dielectric and elastic properties could differ in the respective  systems. A detailed computer simulation may be useful for getting quantitative information.  \\

\section{ Conclusion}
In conclusion, we have measured $\Delta n$,  $\Delta \varepsilon$ and $\Delta K$ of pristine as well as some mixtures of CCH-7 and CCN-47 liquid crystals, possessing identical core structures and antagonistic dipole orientation. Both $\Delta n$  and $\Delta \varepsilon$ of the mixtures decrease  systematically with increasing wt\% of CCN-47.
In particular, at 55wt\%,  $\Delta n$ is reduced by 33\% of the pristine CCH-7 sample and $\Delta \varepsilon$ becomes almost zero. Since the core structures are identical, the large change in the optical and dielectric anisotropies of the mixtures are mostly due to the antagonistic orientation of the dipolar group (-CN). The bend-splay elastic anisotropy $\Delta K$ changes from negative to positive beyond $\sim$37wt\% of CCH-7. The analysis suggests that the antiparallel correlation of dipoles and the resulting molecular association, which is absent in pristine CCN-47 becomes significant beyond 37wt\% of CCH-7 in the mixture.  Thus, the orientation of the strongly polar group (-CN) with respect to the molecular axis influences  the elastic properties significantly, in addition to optical and dielectric properties.\\

{\bf Acknowledgments}:
This work was supported by National Research Foundation of Korea (NRF) 
grant funded by the Korean government (MSIT) (No. 2020R1A2C1006464). SD acknowledges support from DST-FIST-II, School of Physics, University of Hyderabad. \\

\end{document}